\documentclass[runningheads]{llncs}
\usepackage[T1]{fontenc}
\usepackage{graphicx}
\begin{document}
\title{Moving Beyond Review: Applying Language Models to Planning and Translation in Reflection}

\titlerunning{Moving Beyond Review: Applying Language Models in Reflection}

\author{Seyed Parsa Neshaei\inst{1}\orcidID{0000-0002-4794-395X} \and
Richard Lee Davis\inst{2}\orcidID{0000-0002-6175-9200} \and
Tanja Käser\inst{1}\orcidID{0000-0003-0672-0415}}
\authorrunning{S. P. Neshaei et al.}

\institute{EPFL, Lausanne, Switzerland\\
\email{\{seyed.neshaei,tanja.kaeser\}@epfl.ch}
\and
KTH Royal Institute of Technology, Stockholm, Sweden\\
\email{rldavis@kth.se}}
\maketitle              
\begin{abstract}
Reflective writing is known to support the development of students’ metacognitive skills, yet learners often struggle to engage in deep reflection, limiting learning gains. Although large language models (LLMs) have been shown to improve writing skills, their use as conversational agents for reflective writing has produced mixed results and has largely focused on providing feedback on reflective texts, rather than support during planning and organizing. In this paper, inspired by the Cognitive Process Theory of writing (CPT), we propose the first application of LLMs to the \textit{planning} and \textit{translation} steps of reflective writing. We introduce Pensée, a tool to explore the effects of explicit AI support during these stages by scaffolding structured reflection \textit{planning} using a conversational agent, and supporting \textit{translation} by automatically extracting key concepts. We evaluate Pensée in a controlled between-subjects experiment ($N=93$), manipulating AI support across writing phases. Results show significantly greater reflection depth and structural quality when learners receive support during \textit{planning} and \textit{translation} stages of CPT, though these effects reduce in a delayed post-test. Analyses of learner behavior and perceptions further illustrate how CPT-aligned conversational support shapes reflection processes and learner experience, contributing empirical evidence for theory-driven uses of LLMs in AI-supported reflective writing.

\end{abstract}
\section{Introduction}
Reflective writing plays a key role in fostering metacognitive awareness~\cite{perry2019metacognition}. Developing learners' capacity to engage in reflective activities is therefore considered a valuable educational goal across a wide range of contexts, including vocational training~\cite{cattaneo2021reflect,mejia2022evolutionary}. The task of reflective writing involves evaluating learning experiences and adapting them in the form of a written essay. Despite these benefits, learners often struggle to produce high-quality reflections. Meaningful reflection requires more than simple recounting of events; it entails analyzing causes, connecting experiences to prior knowledge, and planning future actions~\cite{middleton2017critical}. Novice learners often struggle with these higher-order processes; they reflect mainly descriptively and not sufficiently analytically~\cite{nehyba2023applications}, resulting in limited learning gains and necessitating reflection support.

Prior work has therefore explored a wide range of approaches to support learners' reflective writing processes. These include instructional scaffolds such as guiding questions~\cite{moussa2015reflective,lai2010examining}, sentence openers~\cite{kingkaew2023learning}, visualizations of writing progress or text structure~\cite{lai2010examining,kingkaew2023learning}, and using structural reflection frameworks~\cite{neshaei2025metacognition,neshaei2025mindmate}. A substantial portion of this research has focused on analyzing and assessing learners' written reflections as a basis for providing adaptive feedback. Such approaches typically identify reflective elements or components within students' texts~\cite{ullmann2019automated,gibson2023reflexive} using traditional machine learning methods, including random forests~\cite{alrashidi2022evaluating,kovanovic2018understand} and topic modeling techniques~\cite{chen2016topic}.

The recent rise of large language models (LLMs) has accelerated this feedback-oriented line of work by enabling more fluent and context-aware analysis of writings~\cite{lee2024design}. Particularly, LLMs have been used to generate adaptive feedback based on rubrics~\cite{awidi2024comparing} or combined with classifiers to personalize feedback on reflection structure~\cite{neshaei2025metacognition}. Beyond text analysis, LLMs also enable conversational agents (CAs) that interact with learners in natural language.
Early CAs for reflective writing relied on pre-scripted dialogues or predefined prompts to guide learners' reflection or to support text structuring~\cite{neshaei2025mindmate}. More recent systems allow free-form interaction with LLM-based CAs. For example, Kumar et al.~\cite{kumar2024supporting} show that access to an LLM supporting self-reflection can improve academic performance, while Kim et al.~\cite{kim2024mindfuldiary} employ a CA to support daily reflective journaling in mental health contexts. Other work uses CAs to guide learners through writing specific components of reflective texts~\cite{neshaei2025metacognition}.

Despite this growing body of work, relatively few studies have examined whether AI-based CAs lead to measurable improvements in reflective writing quality. Evidence regarding the effectiveness of CAs remains mixed: while some studies report positive effects on learners’ perceptions (e.g., perceived usefulness), improvements in reflection depth and analytical quality are  limited or inconsistent~\cite{neshaei2025metacognition}.
A key limitation of existing approaches is that reflection is commonly treated as a single writing-and-feedback activity, overlooking the cognitive processes involved in producing reflective texts. Writing theories, however, emphasize that writing is a complex, multi-stage process. The Cognitive Process Theory of writing (CPT) conceptualizes writing as an interplay of \textit{planning}, \textit{translation}, and \textit{reviewing}~\cite{flower1981cognitive}. Explicitly supporting these stages, which has been shown to be helpful in other writing tasks~\cite{gero2022design,goldi2024intelligent}, might enable deeper and more analytically rich reflective writing.

In this paper, we explore the use of LLMs to explicitly support the \textit{planning} and \textit{translation} stages of reflective writing. Grounded in the Cognitive Process Theory of writing (CPT), we design Pensée, a tool that provides targeted assistance aligned with the cognitive processes underlying reflective writing. Pensée uses an AI-enabled conversational agent that guides learners through structured \textit{planning} activities to foster reflection depth, assists them in \textit{translating} planned elements into coherent reflections through automatic key concept extraction, and supports \textit{reviewing} by providing AI-based feedback on writing structure. We evaluate Pensée in a controlled user study with 93 vocational students, manipulating AI support in the \textit{planning} and \textit{translation} phases. This study addresses three research questions: the effects of providing support in the \textit{planning} and \textit{translation} steps of CPT on \textbf{(RQ1)} the depth and structure of learners' reflective texts, \textbf{(RQ2)} their system usage behavior, and \textbf{(RQ3)} their perceived experience.
Results indicate that AI-supported planning with Pensée improves reflection depth and structural quality, though effects diminish in a delayed post-test. Analyses of learner behavior and perceptions shed light on how such support shapes reflection processes and learner experience. Overall, this work contributes to AIED research by presenting the first CPT-oriented application to explicitly support \textit{planning} and \textit{translation} phases in reflective writing.

\section{Pensée: Reflective Learning with CPT}
\begin{figure}[t]
 \centering
 \includegraphics[width=1\textwidth]{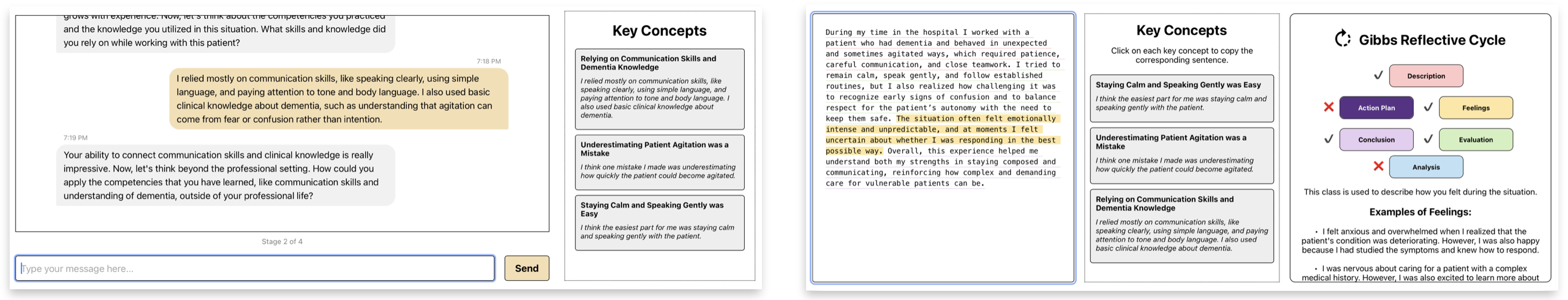}
 \caption{Interface of Pensée consisting of two screens. Left: A CA guides learners in the \textit{planning} phase, with key concepts automatically being extracted to support \textit{translation}. Right: Writing area where learners compose their text using key concepts and receive automated feedback on the text structure during the \textit{review} stage.}
 \label{fig:screenshot-tool}
\end{figure}

To evaluate the effects of AI-based explicit CPT-oriented support on reflective writing, we developed the interactive environment Pensée (see Fig.\ref{fig:screenshot-tool}) that separates reflective writing into the three stages aligned with CPT. Each stage is supported through dedicated interface components and respective AI functionality, enabling students to transform their experiences into coherent and structured reflections. We designed the system to be lightweight and usable in real-world classroom settings without requiring extensive prior training.

\subsection{User Interface Design}
To ground our design, we considered how a student engages in reflective writing according to CPT~\cite{flower1981cognitive}. CPT describes writing as a process involving \textit{planning}, \textit{translation}, and \textit{reviewing}.
The learner typically begins by mentally revisiting a concrete experience through selecting certain moments and \textit{planning} what to include. This step refers to generating and organizing ideas, setting goals, and deciding what to say. They then \textit{translate} these ideas into written text, but might struggle in organizing their thoughts coherently. Finally, they \textit{review} their draft to evaluate whether it fully captures the experience and includes deeper reflection (e.g., analysis of the event) rather than mere description. This step also includes detecting problems and revising content and structure. These three phases of \textit{planning}, \textit{translation}, and \textit{reviewing} indicate the cognitive demands that our tool is designed to scaffold.

As shown in Figure~\ref{fig:screenshot-tool} (left), learners are first presented with the CA interface when logging into the tool. They interact with the AI-powered agent that is designed to support the \textit{planning} step of CPT through asking targeted questions, prompting learners to recall concrete experiences and articulate the learned lessons.
This interface aims to aid students in externalizing and organizing their thoughts before they begin writing their reflection, and makes explicit the kind of idea generation that students would otherwise have to manage internally during their initial planning phase.
To inform the learning design of our CA, we adapted the cycle proposed by~\cite{glogger2012learning} (see Fig.~\ref{fig:chatbot-architecture}-left) to our case. It consists of four main states to cover in order: 1) \textit{metacognition} on talking about why the experience went well or not well, and which competencies they had to use in the event; 2) \textit{connection} on the application of the concepts the students learned in the event outside of their professional life and/or if they remember a prior situation with similar concepts or mistakes; 3) \textit{organization} on the importance of having a clear starting point in reflection, revolving around a main coherent theme, and expanding the text to include prior experiences; and 4) \textit{metacognition again} on asking the learner to revise the conversation, communicate any points missed in their thought process, and planning for changes in their actions in the future.

As the learners progress through the conversation, the ``key concepts'' of their responses are automatically extracted and summarized in a sidebar, with a title and relevant quote from the conversation for each. This sidebar effectively serves as a dynamic planning board at this stage, similar to the notes students might otherwise try to keep mentally or jot down informally, but here persistently organized and visible.
After the end of the conversation session, students are presented with a dedicated writing page (Fig.~\ref{fig:screenshot-tool}-right), in which, corresponding to the \textit{translation} step of CPT, the extracted key concepts remain visible, helping learners convert their planned ideas into a coherent reflective writing without the need to rely solely on memory.
Students can click on any of the key concepts to copy the corresponding quote to the clipboard, ready to paste in the writing area, or rather, read them in a list. The writing page also includes guidance on how to structure a reflection according to the Gibbs reflective cycle, built after prior works on reflection support~\cite{neshaei2025metacognition}; users have the ability to hover their mouse pointer over each component and read the definition of the component plus several examples underneath, which show what well-structured reflections look like beyond simple descriptive texts.
In unsupported writing, this translation phase requires students to transform their loosely organized thoughts into a structured piece, but here, the persistent key concepts reduce that cognitive load and help students maintain alignment in their text with the original experience.

After the essay draft is completed and the learner clicks on the ``Feedback'' button below the typing area, the system supports the \textit{reviewing} step of CPT by providing feedback on which components of the Gibbs reflective cycle are satisfied in the text, underlining the sentences in the text with the color corresponding to each component on the right panel. The checks and crosses on the right panel indicate which components are satisfied and missing in the text, respectively.
Similar to how students reread their drafts to check if they include the necessary components, the tool makes this evaluative process explicit.

As a result, our design explicitly maps each CPT phase to concrete user interactions: the CA for \textit{planning}, the sidebar and extracted key concepts for \textit{translation}, and the writing feedback page for \textit{reviewing}, aligned with how students would otherwise carry out these processes independently.

\vspace{-2mm}
\subsection{AI Models}
\vspace{-1mm}
In Pensée, each CPT stage is supported through different AI mechanisms.

\begin{figure}[t]
 \centering
 \includegraphics[width=0.95\textwidth]{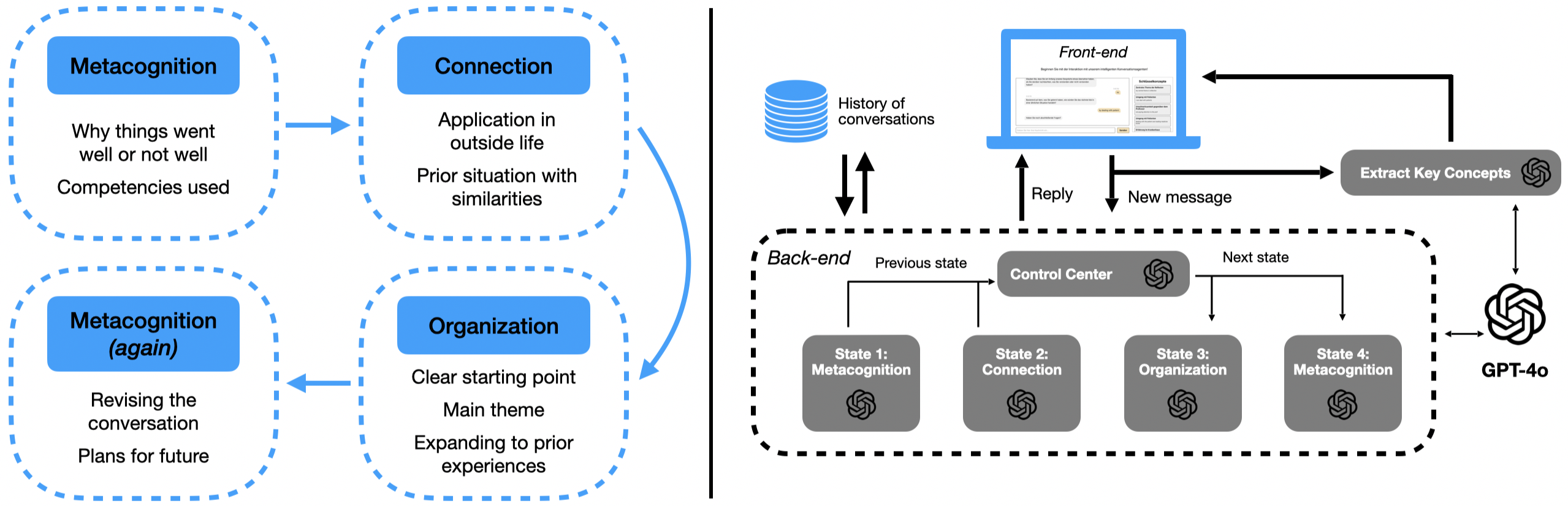}
 \caption{Left: the reflection cycle from Glogger et al.~\cite{glogger2012learning}, starting from top-left, used in designing and steering the model behind our CA. Right: the architecture for the different LLM agents used in the \textit{planning} and \textit{translation} stages in our tool.}
 \label{fig:chatbot-architecture}
\end{figure}

  \noindent  \textbf{Planning.} The planning step is supported through an LLM-based CA, the architecture of which can be seen in Figure~\ref{fig:chatbot-architecture} (right). The CA is implemented as a state machine, progressing through the four main stages of metacognition, elaboration, organization, and again metacognition, as described above\footnote{The full list of questions per state, as well as the types and contents of the prompts used for the LLM, are available in https://github.com/epfl-ml4ed/cpt-reflection}. At each state, another LLM agent (built based on GPT-4o) decides whether all of the questions for the current state have been addressed in the conversation history. If so, the state machine moves to the next state. If not, the LLM will first acknowledge the answer provided by the user, and then continue with asking the remaining questions in the current state. At any point, users can ask follow-up questions, which are handled by the CA and do not trigger a state transition. Upon completion of the fourth state, the conversation ends, and the user is instructed to move to the writing and feedback interface. This model aims to support learners in ``opening up'' and discussing their experiences with probing questions, helping them plan their reflection before writing the final text.

  \noindent \textbf{Translation.} The translation step is supported by an LLM-based agent (built based on GPT-4o) that processes the most recent question–answer pair and generates a corresponding key concept. This concept is displayed in the sidebar and stored in the back-end for later use alongside the writing interface. The agent is prompted to return an automatically-extracted brief title, as well as the relevant quote from the conversation supporting the key concept, only when new information is provided in the learner message. The agent returns structured output as a JSON object containing the key concept title and quote. These key concepts serve as an intermediate representation between planning and writing, helping learners translate ideas surfaced during the conversational planning phase into coherent reflective texts.

  \noindent \textbf{Reviewing.} The reviewing step is supported by an LLM-based agent (GPT-4o) that classifies excerpts of the written reflection according to the six components of the Gibbs reflective cycle~\cite{gibbs1988learning}: Description, Feelings, Evaluation, Analysis, Conclusion, and Action Plan. This structural framework has been used in multiple prior works on reflection (e.g., \cite{ezezika2023development,nurlatifah2023implementation}). The agent is prompted to return a structured output in the form of a JSON list of objects. Each object corresponds to an excerpt of text and contains two fields: 1) the classified component of the Gibbs reflective cycle, and 2) the corresponding text excerpt. The prompt includes a full example of an annotated reflection to guide the model.
We evaluated the classifier on a set of $96$ reflective texts from prior works~\cite{neshaei2025metacognition} and achieved a mean balanced accuracy of 0.66 (Description 0.86, Feelings 0.72, Evaluation 0.50, Analysis 0.37, Conclusion 0.61, Action Plan 0.74), in line with values found in prior works.
In the interface, classification results are displayed both in a dashboard indicating the presence of reflective components (with green checkmarks and red crosses), and directly in the writing area, where text excerpts are highlighted by component. This feedback supports the reviewing process by making the structure of the reflection explicit, which has been shown to improve reflection quality~\cite{neshaei2025metacognition}.

\vspace{-2mm}
\section{Experimental User Study}
\vspace{-2mm}
To evaluate the effectiveness of CPT-oriented AI support for reflective writing, we conducted a controlled study in an authentic classroom setting.

  \noindent \textbf{Conditions.}
We used a fully randomized between-subjects design, comparing two versions of our tool: the main AI-enabled version as the treatment group (TG), and a version without AI support in the \textit{planning} and \textit{translation} steps as the control group (CG).
The CG version of our tool enabled us to isolate the added value of AI support in the \textit{planning} and \textit{translation} steps of CPT. In the CG version, the CA in the \textit{planning} step of CPT is replaced with static text boxes, asking the same reflection questions that the agent asks in the AI-based version, in the same order. For the \textit{translation} step, we directly show the full question and answers of students to the questions. As a result, in this version of our tool, the \textit{interface} still supports students in progressing through \textit{planning}, \textit{translation}, and \textit{reviewing} stages, but without adaptive dialog or automatic extraction of key concepts. We did not isolate the \textit{reviewing} step, and kept AI support in this step for both versions of our tool, building upon prior work that has already shown the benefits of AI support for this step~\cite{neshaei2025metacognition}, removing the necessity to isolate the effects for this stage of CPT.

  \noindent \textbf{Participants.}
The study included  $N=93$ students ($89$ identified as female\footnote{This reflected the distribution of students in the program.}) enrolled in a medical assistance vocational training program in Switzerland (a population comparable to the target groups in prior work \cite{neshaei2025user,neshaei2025metacognition}), ranging in age from 15 to 19 years (mean = $18.47$, SD = $3.11$). Students were randomly assigned to either treatment ($N=45$) or control ($N=48$) groups. All participants consented to the collection of their data, and a parental opt-out procedure was applied for underage students. The study protocol was approved by the university ethics review board (Approval No. HREC 013-2021).

  \noindent \textbf{Procedure.}
\label{ssec:study-procedure}
Our experiment consisted of three sessions: a pre‐intervention (week 1), a learning intervention (week 2), and a delayed (up to 7 days) post‐intervention. 


  \noindent \textit{Pre-intervention Session.}
We started the experiment with a pre-survey, where we collected demographics data and then evaluated the effectiveness of the randomization across conditions using two different constructs of IT usage model (e.g., ``\textit{Using an AI writing assistant to help craft elements of my writing is a good practice}'') and reflective writing knowledge (e.g., ``\textit{I have written reflections before}''), based on the literature~\cite{venkatesh2008technology,neshaei2025metacognition}. Each construct included questions with possible answers from Strongly Disagree to Strongly Agree on a 1-5 Likert scale. To ensure randomization, we used the mean of the results for each question within the constructs.
After the pre-survey, the participants reflected on a situation in the workplace when things did not go as planned. They had to write a minimum of 75 words. This session lasted around 25 minutes per student.

  \noindent \textit{Learning Intervention Session.}
In the learning phase, students used the version of the tool corresponding to their group (treatment or control). They were instructed to use the tool to reflect on a workplace situation that they felt they handled very well. 
After the intervention, they answered a questionnaire, containing questions on a 1-7 Likert scale from the Technology Acceptance Model~\cite{venkatesh2008technology} and used in prior work~\cite{neshaei2025user}, averaged per construct similar to the pre-survey. The survey contained the constructs of excitement after interaction (e.g., ``\textit{Interacting with the tool was exciting}''), perceived usefulness (e.g., ``\textit{Using the tool improves my deep reflective writing performance}''), perceived ease of use (e.g., ``\textit{It is easy for me to become skillful at using the tool}''), technology acceptance (e.g., ``\textit{Assuming the tool is available, the next time I want to write a reflection, I would use it again}''), perceived long-term improvement (e.g., ``\textit{I assume using the tool in the long run will help me improve my abilities to write deeper reflections}''), and correctness (e.g., ``\textit{Adaptive responses, suggestions, and feedback from the tool are correct}'').
This session lasted around 50 minutes per student.

  \noindent \textit{Post-intervention Session.}
Our post-intervention session consisted of four main parts. First, we showed the reflection they had written in the prior session on a page where the students were asked to read it as a knowledge activation task.
Second, as a delayed post-test, we asked a question similar to the question asked in the pre-intervention session (reflecting in a minimum of 75 words on a situation in the workplace when things did not go as planned). During this session, the learners did not receive any CPT-related support from the tool. This session lasted around 35 minutes per student.

\subsection{Measures and Analysis}
We graded each of the three reflections per participant (in the pre-intervention, in the learning session, and in the post-intervention) using two dimensions~\cite{ullmann2019automated}:

  \noindent \textbf{1) Depth.}
We used the reflection strategy posed by~\cite{glogger2012learning}, which was also used to inform the questions asked by the CA, as a basis for our rubric. We graded each reflection on three aspects. For \textbf{metacognition}, we granted one point if the reflection discussed the reasons behind why the experience went well, one point if the reflection discussed the reasons to the contrary (i.e., why it did not go as well), one point for competencies they had to use in the event, and one point for discussing how they would change their behavior or actions in the future. For \textbf{connection}, we granted one point if they discussed if and how they can apply the learned concepts outside of their professional life, and one point if they mentioned another situation with similar competencies or mistakes. Finally, for \textbf{organization}, we granted one point if the reflection had a clear starting point to the problem or the main idea, one point if it consistently revolved around a coherent main theme, and one point if it included proper expansion of the text to past experiences. For the annotation according to this rubric, two researchers labeled ten reflections independently and achieved a Cohen's Kappa of $0.8693$, indicating a \textit{near perfect} agreement. Afterwards, one of the two annotated the rest of the texts on their own.

  \noindent \textbf{2) Structure.}
While we were mainly interested in reflection depth, we also annotated all of the reflections per learner based on the six components of the Gibbs reflective cycle~\cite{gibbs1988learning} as a measure for how well-structured a reflection is. Similar to how it has been done in prior work~\cite{neshaei2025metacognition}, one point was given per each class existing in the text (leading to a minimum of 0 and a maximum of 6 points per text). To annotate the texts, two researchers labeled the Gibbs components independently in five reflections (Cohen's Kappa = $0.9285$, \textit{near perfect} agreement). Afterwards, one of them annotated the rest of the texts on their own.

  \noindent \textbf{Statistical analysis.}
To measure the improvements in reflective writing after providing explicit support, and also to reveal if there can be seen any differences between the AI-based (TG) and non-AI (CG) groups (RQ1), we conducted a series of linear mixed-effects analyses.
We considered one reflective writing outcome variable for structure (the Gibbs score from 0 to 6) and four variables for depth (normalized values of metacognition, connection, organization, and the average of all).
For each outcome variable, we fitted a mixed-effects model with stage (Pre, Tool, Post for the pre-, learning, and post-intervention sessions), study conditions (AI support in TG vs. no AI in CG), and their interaction as fixed effects, and a random intercept for participant. After the global tests of fixed effects, we computed estimated marginal means and conducted planned contrasts with Holm-corrected comparisons. Within each group, contrasts compared performance from Pre to Tool and to Post to evaluate improvement over time. Other contrasts compared TG and CG at each stage to determine whether the presence of AI support led to significant changes in reflective writings.

  \noindent \textbf{Behavioral Analysis.}
To answer RQ2, we focused on the part of the tool different between TG and CG (i.e., the CA versus the static questions). We computed the average number of words per each answer to the bot (in TG) or response to the static questions (in CG) and conducted a statistical comparison with a t-test to see if there are any meaningful differences in the quantity of learner responses. We also particularly computed the mean number of words for the first and last three messages (excluding the initial greeting messages) of the learners, and compared their changes over time using Benjamini-Hochberg corrected t-tests, to find trends in quantitative response behavior. We also analyzed all of the user and CA messages in TG to find evidence of aspects unique to the CA compared to the static interface in CG (that is, the ability of users to ask follow-up questions, and the ability of the CA to engage in free discussion without sticking to asking one question at a time).

  \noindent \textbf{User Experience and Perception Metrics.}
To investigate the learners' perceptions of the intervention and the learning experience using the system (RQ3), we compared responses on our six self-reported perception metrics (see above) between the two study conditions (TG vs. CG).
We first calculated the mean of the responses to the items from each post-survey construct. Then, for each construct, we conducted independent samples Welch t-tests to evaluate the mean differences between the two study groups. We adjusted the p-values using the Benjamini-Hochberg correction. This analysis enabled an analysis of whether perceptions of learning experience differed as a function of receiving AI support.

\vspace{-2mm}
\section{Results}
\label{sec:results}
\vspace{-2mm}
We investigated the effect of CPT-oriented AI-support on the depth and structure of learners' reflections (RQ1), how students interacted with a conversational AI-support system (RQ2), and how they perceived their experience (RQ3).

\begin{figure}[t]
 \centering
 \includegraphics[width=1\textwidth]{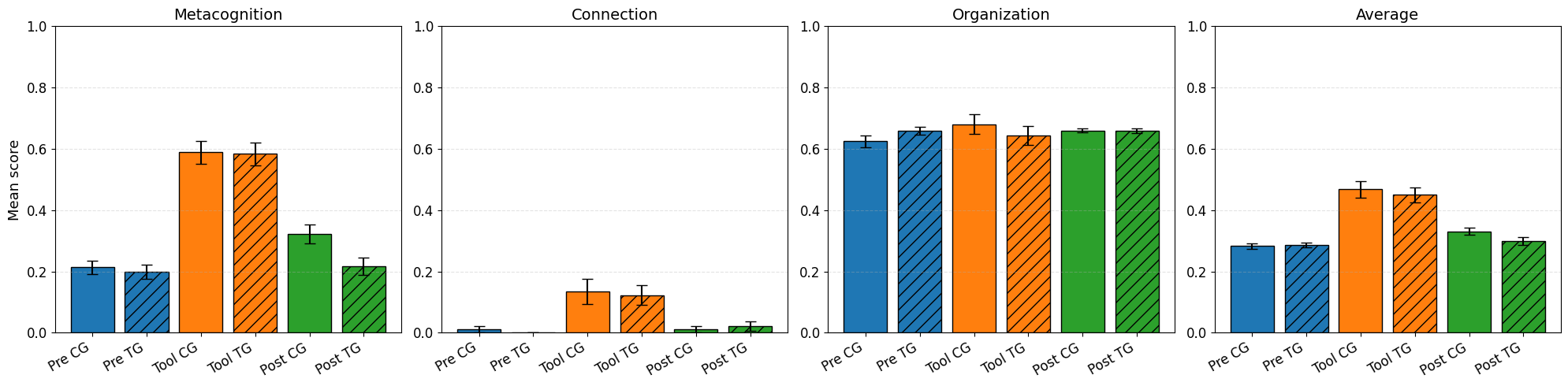}
 \caption{Comparing the depth score of reflections across different stages and groups.}
 \label{fig:depth-results}
\end{figure}

\subsection{RQ1: Effects of Explicit CPT-oriented Support Across Stages}

We analyzed the changes in depth and structure of reflections across stages (Pre, Tool, Post) and groups (TG, CG).

  \noindent \textbf{Depth.}
We observed a significant effect of stage ($p<2e-16$). As can be seen in Fig.~\ref{fig:depth-results}, the reflective writing depth score increased substantially from Pre to Tool (TG: $0.29$ to $0.45$; CG: $0.28$ to $0.47$; both contrasts $p<.0001$), indicating that providing explicit CPT-oriented scaffolding (interface support for planning and translation) improved learners' reflective writing while they used the writing phase of the system, after the end of interaction with the CA. While learners also tended to improve in the delayed post-intervention without support, the difference only showed a trend to significance for CG (TG: $0.29$ to $0.30$; $p=.576$; CG: $0.28$ to $0.33$; $p=.067$), suggesting limited transfer.

  \noindent \textit{Metacognition} scores increased strongly from Pre to Tool in both groups (TG: $0.20$ to $0.58$; CG: $0.21$ to $0.59$; both contrasts $p<.0001$), showing that CPT-oriented support led to deeper metacognitive content in the reflections. From Pre to Post, the improvement was smaller: while both groups had an increased mean, and the effect was significant in CG (from $0.21$ to $0.32$; $p=.016$), it was insignificant in TG (from $0.20$ to $0.22$; $p=.692$). 

  \noindent \textit{Connection} scores increased from Pre to Tool in both groups (TG: $0.00$ to $0.12$; $p=.0005$; CG: $0.01$ to $0.14$; $p=.0003$), but returned to near zero at Post (TG: $0.02$; CG: $0.01$). None of the two groups had a significant change from Pre to Post (TG: $p=.969$; CG: $p=1.000$). This indicates that the intervention led to a higher connection-making when support was present, but the effect did not persist to the delayed post-test without support.

  \noindent \textit{Organization} scores increased from Pre to Tool and Post in CG (Pre: $0.63$; Tool: $0.68$; Post: $0.66$), but slightly decreased from Pre to Tool and then increased back to Post in TG (Pre: $0.66$; Tool: $0.64$; Post: $0.66$). The global stage effect did not show any significant difference ($p=.570$).

\noindent We found no significant effect of condition in any of the statistical tests, not in the overall average depth score nor in any of the rubrics (Overall: $p=.325$; Metacognition: $p=.130$; Connection: $p=.852$; Organization: $p=.964$). Similarly, the interaction effect of stage and condition was also insignificant (Overall: $p=.548$; Metacognition: $p=.163$; Connection: $p=.825$; Organization: $p=.238$).

  \noindent \textbf{Structure.} We observed a significant effect of stage ($p<2e-16$) and an overall clear effect: the structure score increased substantially from Pre to Tool (TG: $2.13$ to $4.38$; CG: $2.10$ to $4.71$; both contrasts $p<.0001$), indicating that Pensée improved the writing structure while they used the system. However, while there was a positive gain to the third, post-intervention session, the difference was not significant (TG: $2.13$ to $2.40$; $p=.409$; CG: $2.10$ to $2.42$; $p=.409$), indicating a limited transfer. Moreover, the effects of condition ($p=.558$) and the interaction of stage and condition ($p=.538$) were not significant.

\noindent \textit{Overall, explicit CPT-oriented support improved reflective writing structure and depth during tool use, especially for metacognition and connection. However, transfer to a delayed post-test without tool support showed limited learning gains. The improvement progress was similar in both groups: they both benefited during tool use, but the additional AI adaptivity and automatic key concepts extraction did not lead to higher scores than static non-AI scaffolding.}

\vspace{-2mm}
\subsection{RQ2: Response Patterns and Interaction Behavior}
Next, we analyzed learners' behavior during the planning phase, which differed between TG and CG. We hypothesized that the CA in TG would encourage longer and deeper interaction compared to the worksheet-style CG interface.

We first examined learners' response length. Students in TG responded to bot messages with a mean of $17.54$ words (SD = $6.35$), while CG students responded to static text boxes with a slightly higher mean of $18.65$ words (SD = $13.24$). However, this difference was not statistically significant ($t=0.505; \; p = .615$). Across time, both groups showed a significant decrease in response length. In TG, responses decreased from $24.50$ words on average (SD = $18.58$) in the first three messages to $8.47$ words (SD = $7.23$) in the last three ($t=8.205; \; p < .0001$). Similarly, in CG, responses decreased from $30.24$ words (SD = $21.73$) to $11.32$ words (SD = $10.86$) ($t=7.832; \; p < .0001$).

A subsequent content analysis of TG responses further showed that only three users asked a single follow-up question each, all of which were clarification questions about terms mentioned by the bot (e.g., ``what are competencies''). This suggests that most users (93\%) interacted with the CA in a way similar to how CG students interacted with static worksheet-style prompts, indicating limited use of the CA's interactive capabilities. Additionally, analysis of the CA messages showed that the CA consistently ended its responses with a question (except for the final congratulatory message), making the interaction structure closely resemble the static question approach used in CG.

\noindent \textit{Overall, we found a notable decrease in response length over time. We also found a high similarity in the usage pattern of the CA in TG and the static questions in CG by the students.}

\vspace{-2mm}
\subsection{RQ3: User Experience and Perception Metrics}
Finally, we explored the responses learners provided to our post-survey questions.

\begin{figure}[t]
 \centering
 \includegraphics[width=0.86\textwidth]{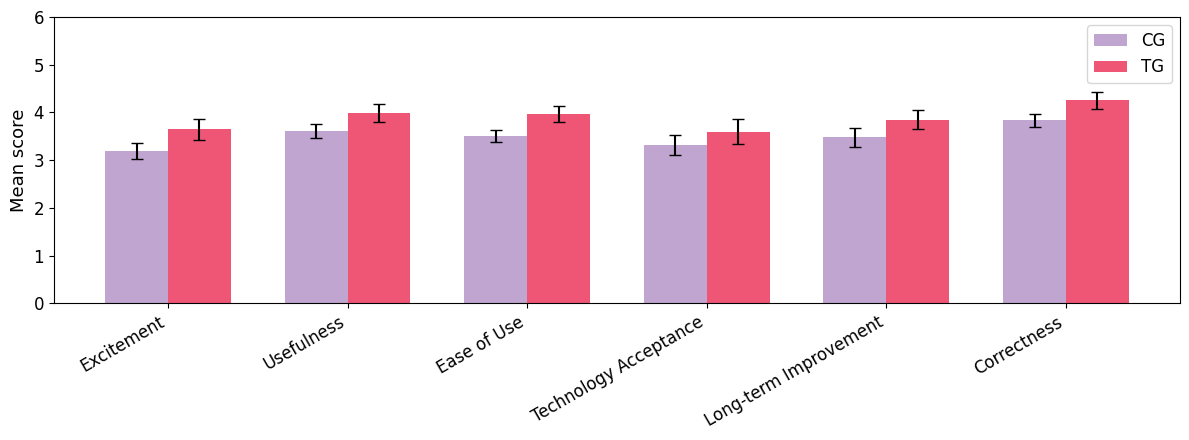}
 \caption{Self-reported perception metrics by group (Mean $\pm$ SEM).}
 \label{fig:perception-metrics}
\end{figure}

As can be seen in Figure~\ref{fig:perception-metrics}, in our post-survey, TG reported higher scores than CG on all metrics, including ease of use (TG: $3.97$; CG: $3.50$), correctness (TG: $4.26$; CG: $3.83$), excitement (TG: $3.64$; CG: $3.20$), usefulness (TG: $3.99$; CG: $3.61$), perceived long-term improvement (TG: $3.84$; CG: $3.48$), and technology acceptance (TG: $3.60$; CG: $3.31$). However, none of these differences remained statistically significant after multiple-comparison Benjamini-Hochberg correction of the p-values ($p > .1$). This indicates that although TG tended to rate the system more positively, the evidence did not show significant perception differences when adding AI support under our current sample size.

\noindent \textit{Overall, AI-supported planning and translation showed consistently more positive perception outcomes than non-AI CPT scaffolding, suggesting a promising advantage that may not have reached significance due to limited statistical power.}

\vspace{-2mm}
\section{Discussion}
\vspace{-2mm}
This study investigated whether grounding support in CPT, particularly targeting \textit{planning} and \textit{translation} stages, would improve learners' reflections and experiences with the system. By explicitly supporting CPT in our tool, we shifted the focus of AI support for reflection from generic feedback on finished texts toward supporting the end-to-end cognitive processes that can also precede writing. Our findings demonstrate that explicit process-oriented scaffolding can meaningfully change reflective writing practice: both study groups showed substantial improvements in reflection depth and structure during tool use, with significant shifts in metacognition, connection-making, and structure.

However, two findings warrant careful consideration. First, the gains observed during tool use did not transfer robustly to the delayed post-test, where learners wrote without scaffolding. This suggests that short-term exposure, while boosting performance, may not produce lasting changes in reflective writing skills (a common challenge in writing interventions). Second, the AI-supported condition showed no significant advantage over the static scaffolding condition. Both groups improved similarly, indicating that the primary driver was the explicit CPT support itself, rather than AI adaptivity or conversational interaction.

  \noindent \textbf{Implications and Next Steps.} These findings carry implications for the design, study, and evaluation of LLM-based learning tools.

 \noindent \textit{Design.} Our behavioral analysis revealed that students in the AI condition largely treated the CA like a static worksheet. Response patterns were similar across conditions, response length declined sharply over turns, and only 7\% of users ever asked a follow-up question. This implies that the CA's design may have inadvertently replicated the static scaffolding experience. To realize the interactive potential of LLMs, future designs should explicitly scaffold richer interaction patterns (e.g., pre-defined buttons for asking follow-ups, giving examples, or challenging reasoning). Changes like these, as well as considering careful evaluation of the LLM outputs, may be necessary to unlock the potential of CAs for learning.

  \noindent \textit{Long-term impact.} Although not statistically significant, the AI-supported group consistently rated the system higher on ease of use, excitement, usefulness, and willingness to reuse, matching prior CA and writing works (e.g., \cite{wambsganss2021arguetutor}). If learners prefer AI-supported tools, impact may emerge through sustained engagement, with higher adoption leading to more practice over time. This suggests a shift in research focus from single-session outcomes toward understanding how perception advantages translate into long-term tool use and, ultimately, learning gains. Designing for repeated engagement rather than one-off interventions may be where AI support proves invaluable.

\noindent \textit{Evaluation.} The lack of significant learning gains reflects a common issue in writing interventions: short exposure to support can boost performance, but may not lead to transfer performance once the scaffolds are removed. This points to a methodological gap the community should address: many reflection or LLM-writing studies evaluate same-session outcomes (e.g., \cite{neshaei2025metacognition,mejia2025enhancing}), while delayed post-tests are less common, and set a higher bar for performance. This gap may lead the field to overestimate intervention effects. Future work should adopt delayed post-tests as standard practice and design longitudinal studies that capture learning trajectories over time to distinguish between performance and long-term learning.


  \noindent \textbf{Conclusion.} We presented Pensée, the first application of LLMs to explicitly support the planning and translation stages of reflective writing, grounded in CPT. Our results show that process-oriented scaffolding significantly improves reflection depth and structure during use. Yet the absence of significant differences between AI-supported and static conditions offers a grounding insight: LLMs are powerful tools, but they are not magic. Realizing their potential for learning requires careful, theory-driven design and rigorous evaluation. This work identifies concrete directions for the field: engineering interfaces that unlock genuine conversational engagement, investigating whether perception advantages compound into learning gains through sustained use, and adopting evaluation practices that measure lasting impact.

\begin{credits}
\subsubsection{\ackname}
This project was substantially financed by the Swiss State Secretariat for Education, Research, and Innovation (SERI). We thank the teachers who enabled us run our study, as well as Dominik Glandorf, Ekaterina Shved, and Peter Bühlmann for their assistance with data collection, Fatma Betül Güres for proofreading the manuscript, and Roman Rietsche and Thiemo Wambsganss for their contributions to ideation and providing valuable feedback.

\subsubsection{\discintname}
The authors have no competing interests to declare that are relevant to the content of this article.
\end{credits}

%
%

\bibliographystyle{splncs04}
\bibliography{reflectiontool}

%

\end{document}